# The MICADO first light imager for the ELT: overview and current status

E. Sturm[a], R. Davies[a], J. Alves[b], Y. Clénet[c], J. Kotilainen[w], A. Monna[d], H. Nicklas[e], J.-U. Pott[f], E. Tolstoy[h], B. Vulcani[g], J. Achren[w], S. Annadevara[d], H. Anwand-Heerwart[e], C. Arcidiacono[g], S. Barboza[f], L. Barl[a], P. Baudoz[c], R. Bender[a,d], N. Bezawada[i], F. Biondi[a], P. Bizenberger[f], A. Blin[s], A. Boné[f], P. Bonifacio[k], B. Borgo[c], J. van den Born[n], T. Buey[c], Y. Cao[a], F. Chapron[c], G. Chauvin[v], F. Chemla[k], K. Cloiseau[c], M. Cohen[k], C. Collin[c], O. Czoske[b,o], J.-O. Dette[e], M. Deysenroth[a], E. Dijkstra[n], S. Dreizler[e], O. Dupuis[c], G. van Egmond[n], F. Eisenhauer[a], E. Elswijk[n], A. Emslander[a], M. Fabricius[a], G. Fasola[k], F. Ferreira[c], N. M. Förster Schreiber[a], A. Fontana[y], J. Gaudemard[k], N. Gautherot[t], E. Gendron[c], C. Gennet[c], R. Genzel[a], L. Ghouchou[c], S. Gillessen[a], D. Gratadour[c], A. Grazian[g], F. Grupp[a,d], S. Guieu[r], M. Gullieuszik[g], M. de Haan[n], J. Hartke[w], M. Hartl[a], F. Haussmann[a], T. Helin[z], H.-J. Hess[d], R. Hofferbert[f], H. Huber[a], E. Huby[c], J.-M. Huet[k], D. Ives[i], A. Janssen[n], P. Jaufmann[n] , T. Jilg[a], D. Jodlbauer[m], J. Jost[f], W. Kausch[o], H. Kellermann[d], F. Kerber[i], H. Kravcar[d], K. Kravchenko[a], C. Kulcsár[u], H. Kuncarayakti[w], P. Kunst[n], S. Kwast[n], F. Lang[a], J. Lange[a,d], V. Lapeyere[c], B. Le Ruyet[c], K. Leschinski[b], H. Locatelli[t], D. Massari[h], S. Mattila[w], S. Mei[p], F. Merlin[c], E. Meyer[t], C. Michel[l], L. Mohr[f], M. Montargès[c], F. Müller[f], N. Münch[f], R. Navarro[n], U. Neumann[f], N. Neumayer[f], L. Neumeier[a], F. Pedichini[y], A. Pflüger[a], R. Piazzesi[y], L. Pinard[l], J. Porras[c], E. Portaluri[x], N. Przybilla[o], S. Rabien[a], J. Raffard[c], R. Ragazzoni[g], R. Ramlau[m,q], J. Ramos[f], S. Ramsay[i], H.-F. Raynaud[u], P. Rhode[e], A. Richter[e], H.-W. Rix[f], M. Rodenhuis[j], R.-R. Rohloff[f], R. Romp[n], P. Rousselot[t], N. Sabha[o], B. Sassolas[l], J. Schlichter[d], M. Schuil[n], M. Schweitzer[i], U. Seemann[i], A. Sevin[c], M. Simioni[g], L. Spallek[a], A. Sönmez[a], J. Suuronen[z], S. Taburet[k], J. Thomas[a], E. Tisserand[t], P. Vaccari[x], E. Valenti[i], G. Verdoes Kleijn[h], M. Verdugo[b], F. Vidal[c], R. Wagner[q], M. Wegner[d], D. van Winden[n], J. Witschel[e], A. Zanella[g], W. Zeilinger[b], J. Ziegleder[a], and B. Ziegler[b]

*[a]Max Planck Institute for extraterrestrial Physics, 85748 Garching, Germany; [b]Institute for Astrophysics, University of Vienna, 1180 Wien, Austria;[c]LESIA, Observatoire de Paris, Université PSL, CNRS, Sorbonne Université, Université Paris Cité 92195, France; [d]University Observatory of LMU Munich, 81679 München, Germany; [e]Institute for Astrophysics and Geophysics, Georg-August-Universität Göttingen, 37077 Göttingen, Germany; [f]Max Planck Institute for Astronomy, 69117 Heidelberg, Germany; [g]INAF-OaPD, 35122 Padova PD, Italy; [h]Kapteyn Astronomical Institute, 9700 AV Groningen, The Netherlands; [i]European Southern Observatory, 85748, Garching, Germany; [l]LMA, IP2I Lyon, CNRS, Université Claude Bernard, 69622 Villeurbanne Cedex; France; [j]Leiden Observatory, University of Leiden, 2300 RA Leiden, The Netherlands; [k]GEPI, Observatoire de Paris, Université PSL, CNRS, 92195, Meudon, France; [m]Industrial Mathematics Institute, Johannes Kepler University Linz, 4040 Linz, Austria; [n]NOVA-ASTRON, Oude Hoogeveensedijk, 7991 PD Dwingeloo, The Netherlands; [o]Institut für Astro- und Teilchenphysik, Universität Innsbruck, 6020 Innsbruck, Austria; [p]Université Paris Cité, CNRS(/IN2P3), Astroparticule et Cosmologie, 75205 Paris cedex 13, France; [q]Johann Radon Institute for Computational and Applied Mathematics (RICAM), 4040 Linz, Austria; [r]EFISOFT, IPAG, Univ. Grenobles Alpes, CNRS, 38400 Saint-Martin-d'Hères, France; [s]Division Technique INSU, CNRS, 91190 Orsay, France; [t]OSU THETA, Univ. de Franche-Comté, CNRS, 25000 Besançon, France; [u]LCF, Université Paris-Saclay, Institut d'Optique Graduate School, CNRS, Laboratoire Charles Fabry, 91127, Palaiseau, France; [v]Lagrange, OCA, CNRS, Université de la Côte d'Azur, 06304 Nice, France; [w]University of Turku, FINCA, 20014 Turku, Finland; [x]INAF-OaAB, 64100 Teramo, Italy; [y]INAF-OaR, 00036, Rome, Italy; [z]LUT University, 53850 Lappeenranta, Finland*

*sturm@mpe.mpg.de, phone: +49-89-30000-3806

## ABSTRACT

MICADO is a first light instrument for the Extremely Large Telescope (ELT), set to start operating later this decade. It will provide diffraction limited imaging, astrometry, high contrast imaging, and long slit spectroscopy at near-infrared wavelengths. During the initial phase operations, adaptive optics (AO) correction will be provided by its own natural guide star wavefront sensor. In its final configuration, that AO system will be retained and complemented by the laser guide star multi-conjugate adaptive optics module MORFEO (formerly known as MAORY). Among many other things, MICADO will study exoplanets, distant galaxies and stars, and investigate black holes, such as Sagittarius A* at the centre of the Milky Way.

After their final design phase, most components of MICADO have moved on to the manufacturing and assembly phase. Here we summarize the final design of the instrument and provide an overview about its current manufacturing status and the timeline. Some lessons learned from the final design review process will be presented in order to help future instrumentation projects to cope with the challenges arising from the substantial differences between projects for 8-10m class telescopes (e.g. ESO’s VLT) and the next generation Extremely Large Telescopes (e.g. ESO’s ELT).

Finally, MICADO's expected performance will be discussed in the context of the current landscape of astronomical observatories and instruments. For instance, MICADO will have similar sensitivity as the James Webb Space Telescope (JWST), but with six times the spatial resolution.

## 1. INTRODUCTION

MICADO[1], the Multi-AO Imaging Camera for Deep Observations, will equip the ELT with a first light diffraction limited imaging capability at near-infrared wavelengths. The instrument will work with a multi-conjugate laser guide star adaptive optics system (MCAO, developed by the MORFEO consortium[2]) as well as a single-conjugate natural guide star adaptive optics system (SCAO, developed by the MICADO consortium). It will interface to the MORFEO warm optical relay that re-images the telescope focus. In this configuration, both MORFEO and SCAO are available. In an initial phase, MICADO will also be able to operate with just the SCAO system in a ‘stand-alone’ mode, using a simpler optical relay that interfaces directly to the telescope.

MICADO will address a large number of science topics that span the key elements of modern astrophysics. The science drivers focus on several main themes: the dynamics of dense stellar systems, black holes in galaxies and the centre of the Milky Way, the star formation history of galaxies through resolved stellar populations, the formation and evolution of galaxies in the early universe, planets and planet formation, and the solar system. To address these, MICADO will exploit its key capabilities of sensitivity and resolution, which are in turn leveraged by its observing modes of imaging, astrometry, coronography, and spectroscopy.

## 2. DESIGN

MICADO provides a high- (HRI) and low- (LRI) resolution imaging mode with an unvignetted field of view (FOV) of 19“x19” at a pixel scale of 1.5 mas, and 50.5”x50.5” at a pixel scale of 4 mas, respectively. MICADO comprises the infrared focal plane imager with its 3x3 array of 4Kx4K HgCdTe (HAWAII-4RG) detectors and a compact cross-dispersing slit spectrometer and will operate in the wavelength range of 0.8 to 2.4 µm. High contrast imaging is enabled via either focal plane (classical Lyot coronagraphs) or pupil plane masks (vector-Apodizing Phase Plate (vAPP) & sparse aperture masks (SAM)). In the early phase operation of MICADO at ELT (so-called ’stand-alone’ configuration), MICADO will benefit from a SCAO correction that makes use of the ELT M4 and M5 adaptive mirrors.

This mode will allow for diffraction limited observation in a smaller (but not necessarily central) FOV making use of wave front sensing of a nearby single natural guide star. In this phase the MICADO instrument will be hosted on the ELT Nasmyth platform A, utilizing the optical straight through port of the ELT “Pre-Focal Station”. In its final configuration MICADO will be mounted on Nasmyth platform B, together with MORFEO.

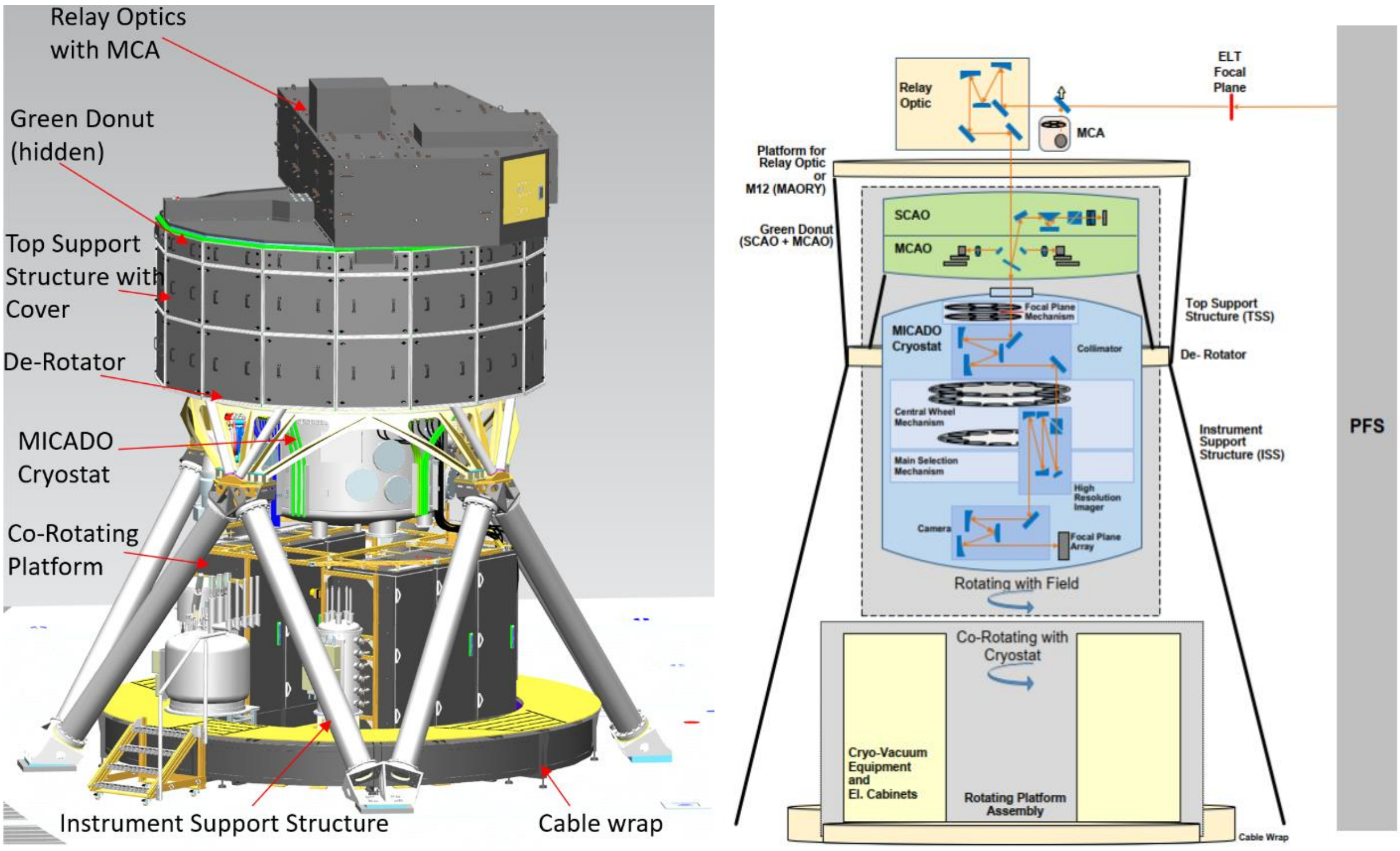

Figure 1: Global architecture of MICADO in the stand-alone mode on Nasmyth platform A *(left)*, a rendering of the instrument, with the prominent components labelled. *Right*: Functional view of MICADO in stand-alone configuration.

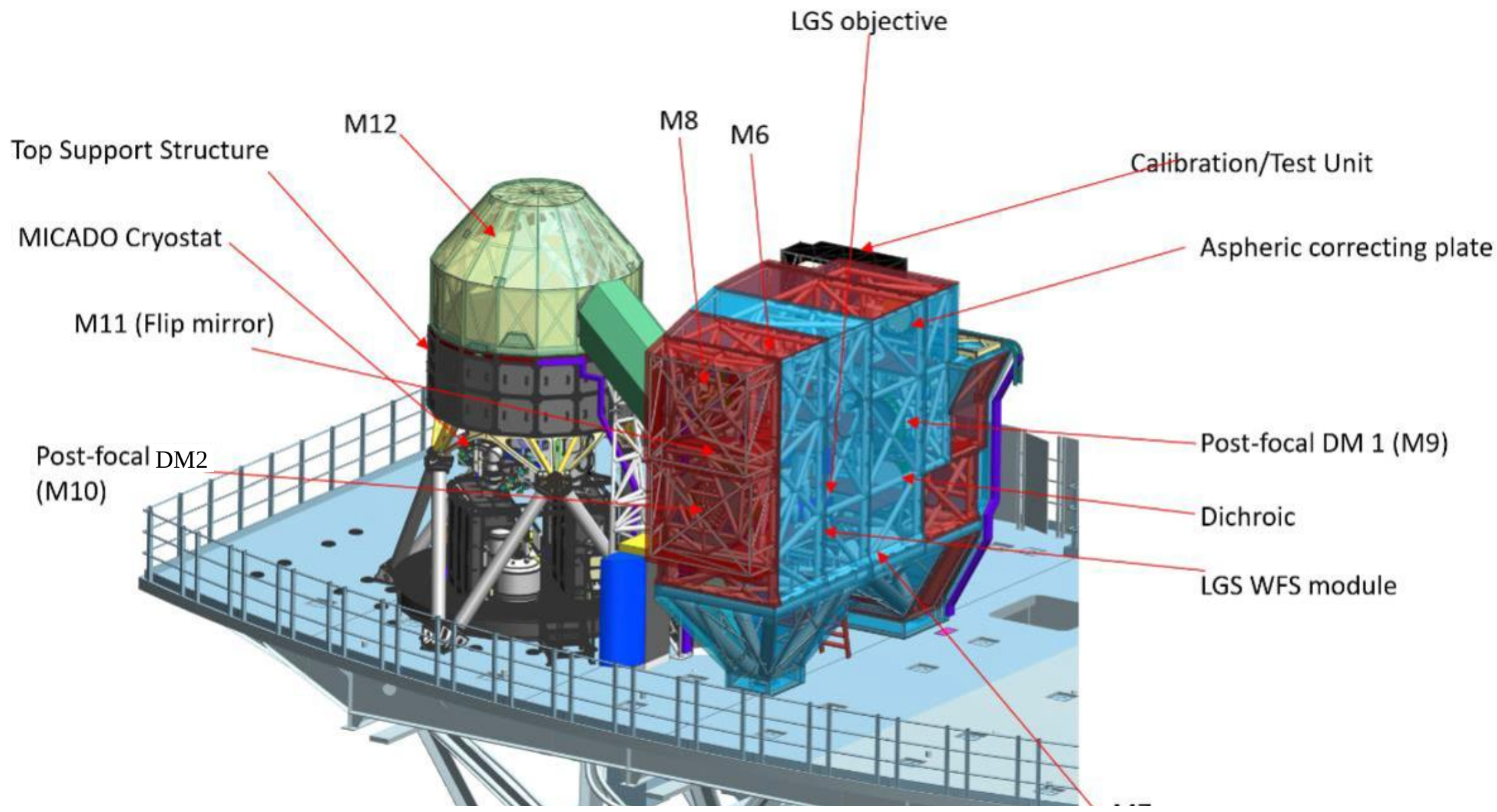

Figure 2: MICADO in its final configuration with MORFEO on Nasmyth platform B (shown without its service platform). The last MORFEO fold mirror M12, mounted on the top platform of MICADO, replaces the stand-alone relay optics. In this configuration, the calibration unit is moved to the calibration unit selector on the MORFEO bench.

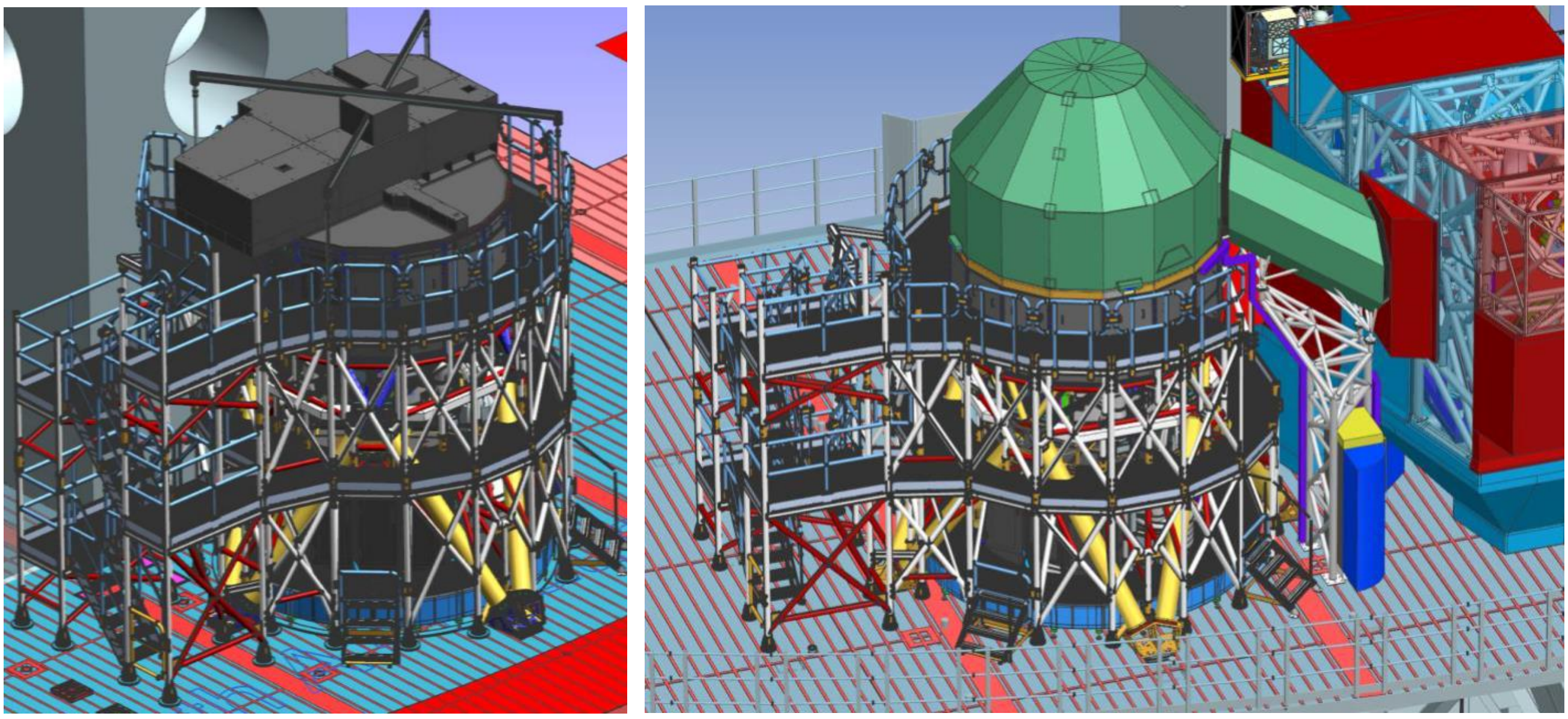

Figure 3: MICADO and its access structure in stand-alone mode on Nasmyth platform A (left), and in its final configuration with MORFEO on Nasmyth platform B (right).

The global architecture and a functional view of the MICADO instrument in its MICADO/SCAO 'stand-alone' configuration are given in Figure 1. Figure 2 shows the final configuration with MORFEO.

The MICADO instrument comprises the following top-level subsystems:

- The *Calibration assembly (MCA)*[3], which enables the calibration of the science instrument, by creating suitable sources in a focal plane matching that of the ELT. It consists of 3 units: The Astrometric calibration unit (ACU), the Movable source calibration unit (MCU), and the Flat-field and wavelength calibration unit (FCU).
- The *Relay optics (RO)*[4][5] system, which transfers the ELT focal plane into the instrument. This will be used only in the stand-alone mode. With MORFEO, the system will be replaced by a single-fold mirror.
- Two *NGS WFS module*[6][7] (SCAO and MORFEO's Low Order and Reference module LOR), containing natural guide star wavefront sensors for the adaptive-optics systems. They are accommodated in the so-called *Green Doughnut (GD)* envelope located on top of the cryostat as close as possible to the entrance focal plane of the MICADO cryostat. SCAO hosts its own calibration unit SCU. SCAO and LOR are connected with independent rigid and stiff mechanical interface structures to the rotating central cryostat flange to provide a very stable rotating system (see below). This rotating system is designed such that its horizontal location of the centre of gravity is close to the rotation plane of the De-Rotator to avoid torque effects.
- The *De-Rotator (DR),* which rotates the entire cryostat and the NGS WFS modules to match the orientation of the sky as seen by the instrument (i.e. de-rotating the field and pupil).
- The *Cryostat*[8], which is the core of MICADO, a liquid nitrogen continuous flow cryostat which cools its internal subsystems (cold optics and mechanics, detector arrays) to ~80 K.
- and contains the cold optics system and detector array[15].
- The *Support structure (ST0)*[9], which keeps the various components at their necessary locations. The support structure consists of two parts:

  The *Instrument Support Structure (ISS)* and the *Top Support Structure (TSS)*. The highest part of the system is mounted on the TSS. This is a non-rotating optical platform on top, which can accommodate either the RO and the MCA in the MICADO/SCAO 'stand-alone' configuration, or MORFEO's last fold mirror M12, including a cover in MICADO/MORFEO (M&M) configuration. The ISS is a Serrurier like truss structure which mechanically supports the de-rotator, which in turn supports and rotates the cryo-vacuum vessel.

- The *Rotating Platform Assembly (RPA)*, including the cable wrap, which allows much of the electronics to co-rotate with the instrument, keeping cable lengths short; the cable wrap enables access to external electronics and service supplies. On this platform auxiliary cryo-vacuum components, electrical cabinets with a water-cooling system and the detector readout electronic are placed. The cable wrap around the co-rotating platform provides a ± 270° moving duct for the harness and the piping system.
- The *Instrument Electronics and control system*, which facilitate control and monitoring signals, and also provide power for all devices.
- The *Control software*, which supports observation preparation and execution, coordinating the operation of the instrument with the telescope and adaptive-optics systems.
- The *Pipeline software*, which performs data processing for a quick look at the observatory, for archive and trending analyses, and for the user back home. It includes a PSF reconstruction module[16],[17] and a data simulator.
- Handling, maintenance and AIT equipment
- A service platform ("*access structure*"), see Figure 3.

## 3. STATUS

At the time of writing this article, most parts of MICADO are already in their manufacturing, assembly, integration and test phase (MAIT), while a few last aspects of the preceding phase, the final design review (FDR), are being finalized. ESO and the consortium expect this last part of the FDR process to be concluded in July 2024. The philosophy of and lessons learned from this hybrid approach (two overlapping phases) are explained in the next section. The remaining aspects under review are the electronics and the service platform ("access structure"), as both the (control) electronics design and the access structure design could only be finalized once the design of most of the other sub-systems was frozen. In addition, ESO's standards and requirements for the electronics schematics had changed (see §6) and a re-formatting became necessary.

A few examples of hardware parts that are well advanced in their manufacturing process are shown in Figure 4: the heart of the instrument, the cryostat, is being assembled right now and will be ready for system integration by the end of 2024. It recently passed successfully a critical vacuum test. The De-Rotator with its conical support structure, its adapter ring and roller bearings, has been assembled and is now ready for sub-system testing. Its mating with the cryostat is one of the first steps of the system integration which will start at the end of 2024. The optical bench of the Relay Optics was completed early in 2024. It has been blackened for straylight reduction and is now being equipped and tested with optical components such as the mirrors of the pre-optics and the calibration units in a dedicated assembly hall. The pupil wheel, one of the cold mechanisms, with its 18 filter positions for pupil masks (Lyot stops, aperture masks, vAPP) and narrow band filters has undergone very successful cold tests in April and May this year. Altogether the MICADO cryostat will host three filter wheels. Many of the numerous optical and mechanical SCAO parts, like the K-mirror, are also already in house.

The schedule for the MAIT phase is divided into two parts. The first part takes place in the main integration hall at MPE, Garching. It comprises mostly the cold system (cryostat, cold optics, cold mechanisms, detectors), the De-Rotator, and a dedicated support structure for that phase. It ends with the delivery of the SCAO system.

The final integration and end-to-end test phase for the MICADO system (including then also all warm components like Relay Optics, MCA, Co-Rotator, the real Support structure and the Access Structure) will take place in ESO's Large Integration Hall in Garching, Germany. MICADO will be packed and shipped to Chile directly from there. Currently Preliminary Acceptance Europe (PAE, the last milestone before shipping) is expected to happen in 2028.

This two-phased approach for the MAIT schedule is an example of the many technical, logistical and managerial challenges that arise from Telescopes and their instruments becoming ever bigger, as both the height of and weight (crane load) exceed the allowed limits in the main MICADO system integration hall at MPE. To accommodate even the first MAIT phase (with reduced height and weight) it was still necessary to cut a hole in the roof of the hall (with a removable cover) to allow for lifting of sub-units with the help of an external crane for heavy loads.

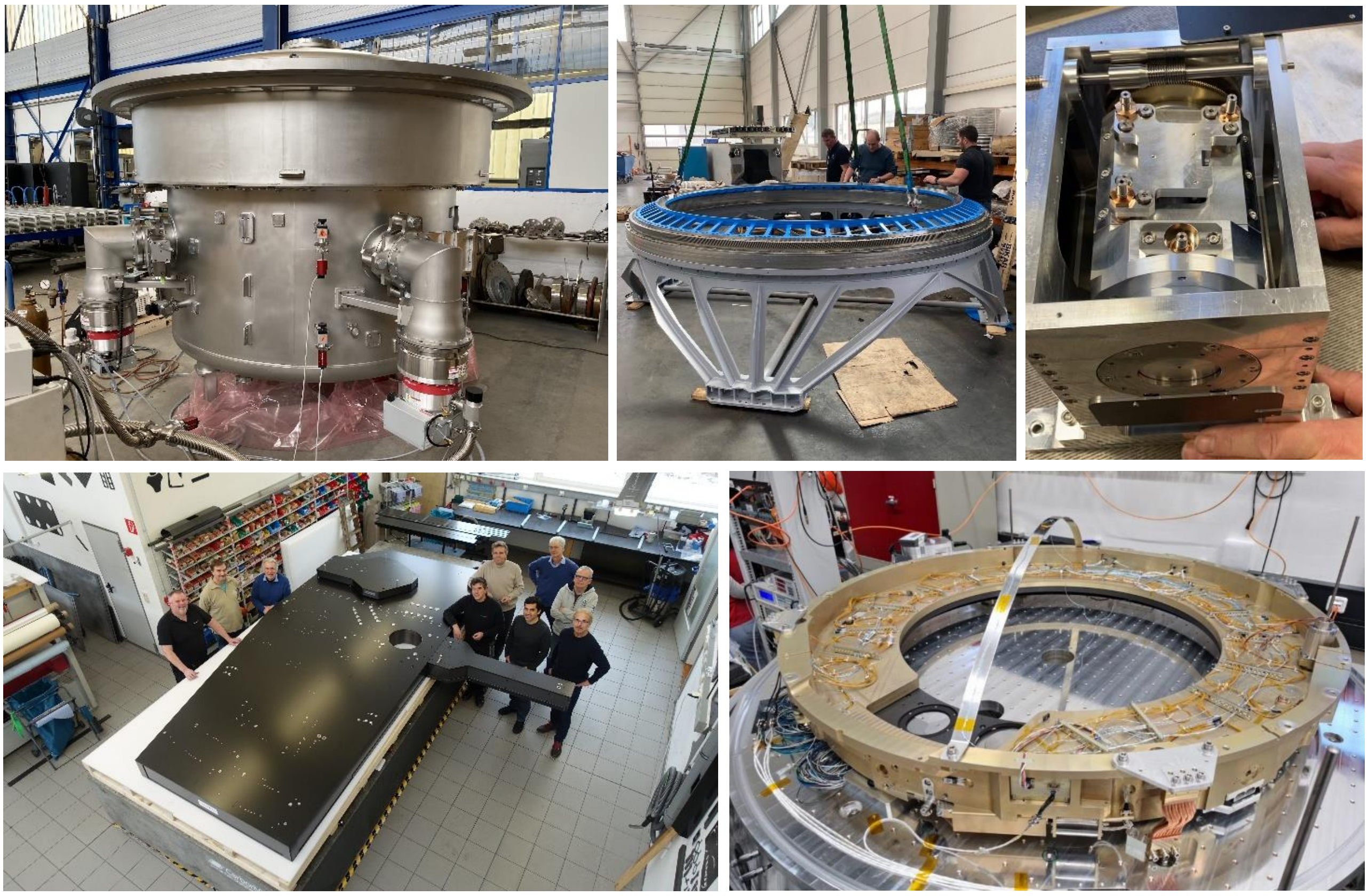

Figure 4: *Top left:* the MICADO cryostat during (successful) vacuum testing. *Top middle:* (successful) acceptance testing of the De-Rotator. *Top right:* The SCAO K-mirror. *Bottom left:* The MICADO Relay Optics Bench during (successful) acceptance by the MPIA MICADO team. *Bottom right:* The pupil wheel mechanism being prepared for the (successful) cold sub-system tests.

## 4. THE FINAL DESIGN REVIEW APPROACH

As described above, at the time of writing this article, MICADO is in a hybrid phase, with most parts in their MAIT phase while finishing up the last aspects of the FDR process. This transition with two overlapping project phases is the effect of the staggered FDR approach that we pursued; it was in fact intended. In order to make the review of such a big and complex project more manageable, and in order to ensure that the procurement and manufacturing of ready-to-go items, and in particular long lead items, was not delayed by other sub-units that needed a little more time to finalize their design, ESO and the MICADO Consortium together decided to split the FDR process in (initially 3, then) 4 sub-reviews.

Among the first items undergoing FDR were the MAIT plans for each sub-unit and the system level, management aspect (risks, hazards, schedules, etc.), operational concepts, the RAMS[10] approach, detectors, and (parts of) the pipeline software. The detailed sub-system designs were then subject to the subsequent sub-reviews.

This approach has indeed helped to keep the manufacturing phase of MICADO within a tight schedule. But it came, of course, with many risks. For instance, all internal interfaces needed to be frozen at the very beginning of the process (even if some units were still being designed), and the full and up-to-date system view would only come at the end of the process.

If such a process stretches over too long a time interval, there is a danger of losing the big picture, of getting lost in low level details and a nice-to-have wish list mindset. It became clear that a very collaborative approach, rather than an over-simplified customer-seller picture between ESO and the instrument consortium (which may be more appropriate for industrial contracts), was needed to lead this process to a successful conclusion. The cryo-control software design - with ESO providing templates and libraries for their preferred HTML5 system - and an elaborate and sophisticated global vibration analysis of MICADO on the Nasmyth platform - performed by ESO with input from the consortium – are two positive examples of this collaborative approach.

A good summary of some of the challenging aspects of the review process can be found in ESO's article about lessons learned from recent major projects in the recent messenger issue #192[11].

## 5. SOME CHALLENGES

Besides the challenges posed by the FDR process of complex projects, there are of course all kinds of other challenges, be it of technical or managerial character. The many technological innovations that were required to fulfill the demanding technical and top-level specifications shall be described in a forthcoming publication. Here we want to mention some of the more managerial aspects.

The ELT and its first-generation instruments are, in almost every aspect, very comparable to space projects. Be it the long timescales, the simultaneous development of telescope and instruments (with the need to constantly iterate on interfaces, requirements and design), the large consortia, the high costs, or the demanding maintenance aspects, to mention only a few. One example that illustrates the difference with projects for currently existing telescopes is the sheer size and mass of the instruments. MICADO has a mass of more than 20 tons (the equivalent of four adult male elephants) and is more than 6 meters high (like a giraffe). This has consequences, for instance, for the necessary size of integration halls and the need for heavy load cranes (see above), for the transport of oversized containers (be it on the road or on aircrafts), for maintenance activities (the length of a standard-arm may not be long enough to reach in to adjust components), or other aspects of the complex logistics for assembly, integration and testing. All going far beyond the needs for the development of instruments on 10m-class telescopes.

In contrast to most space projects repair missions are possible, but they would be extremely laborious and lead to long down times. In addition, ESO intends to operate the Paranal and Armazones sites as single (multi-site) observatory. Among other reasons this may be dictated by the limited resources. In any case, both aspects (difficult-to-repair and to operate) require increased levels of automation and more complex development and testing, in order to minimize on-site activities and commuting. The focus on RAMS (Reliability, Availability, Maintainability, Safety), on automation, and on self-diagnosis (e.g. for cryo-control) is considerably increased compared to past projects, once more very comparable to space missions. For the maintenance strategy this requires better scheduling of maintenance activities and an increased, condition-based/predictive maintenance plan to minimize the need for corrective/preventive maintenance activities.

The large size of the consortia causes severe management overheads and leads to much increased communication needs. It also leads to more multi-tasking, as large parts of the community are involved in several projects. One particular example of the need for good communication and collaboration across consortia is the interface between MICADO and MORFEO. This interface is of course key for the performance of the instrument. It is formally "owned" by ESO, i.e. there are three parties involved (ESO, MICADO, MORFEO). Due to the different development schedules of MICADO (post FDR) and MORFEO (pre-FDR), managing this interface is a major challenge. It requires many working meetings in a collaborative spirit, and an early freeze of essential interfaces or at least agreed parameter envelopes to have a fully consistent interface in place.

The long duration of modern ground-based telescope and instrumentation projects is a major problem for any personnel management. It is at odds with national regulations for fixed term contracts and makes it difficult to retain the staff. The duration is also at odds with national funding cycles (which are usually of the order of a few years only). As long as our current and future large telescope projects are demanding as space projects but funded at the traditional lower level of previous ground-based projects this mismatch will put future instrument development at a severe risk.

## 6. PERFORMANCE IN CONTEXT

MICADO will address a large number of science topics that span the key elements of modern astrophysics. The science drivers[5] focus on several main themes: the dynamics of dense stellar systems, dark matter and IMBHs; super-massive black holes in galaxies and the centre of the Milky Way; the star formation history of galaxies through resolved stellar populations, the formation and evolution of galaxies in the early universe, planets and planet formation, and the solar system. The main science cases and astrophysical applications are summarized in Davies et al. 2018[1].

Here we want to put the performance of ELT/MICADO in the context of the current landscape of astronomical observatories and instrumentation. MICADO at the ELT will be very complementary to the JWST. It reaches similar sensitivity but at 6-7 times higher linear spatial and spectral resolution (see Figure 5; Figure 6; Figure 7).

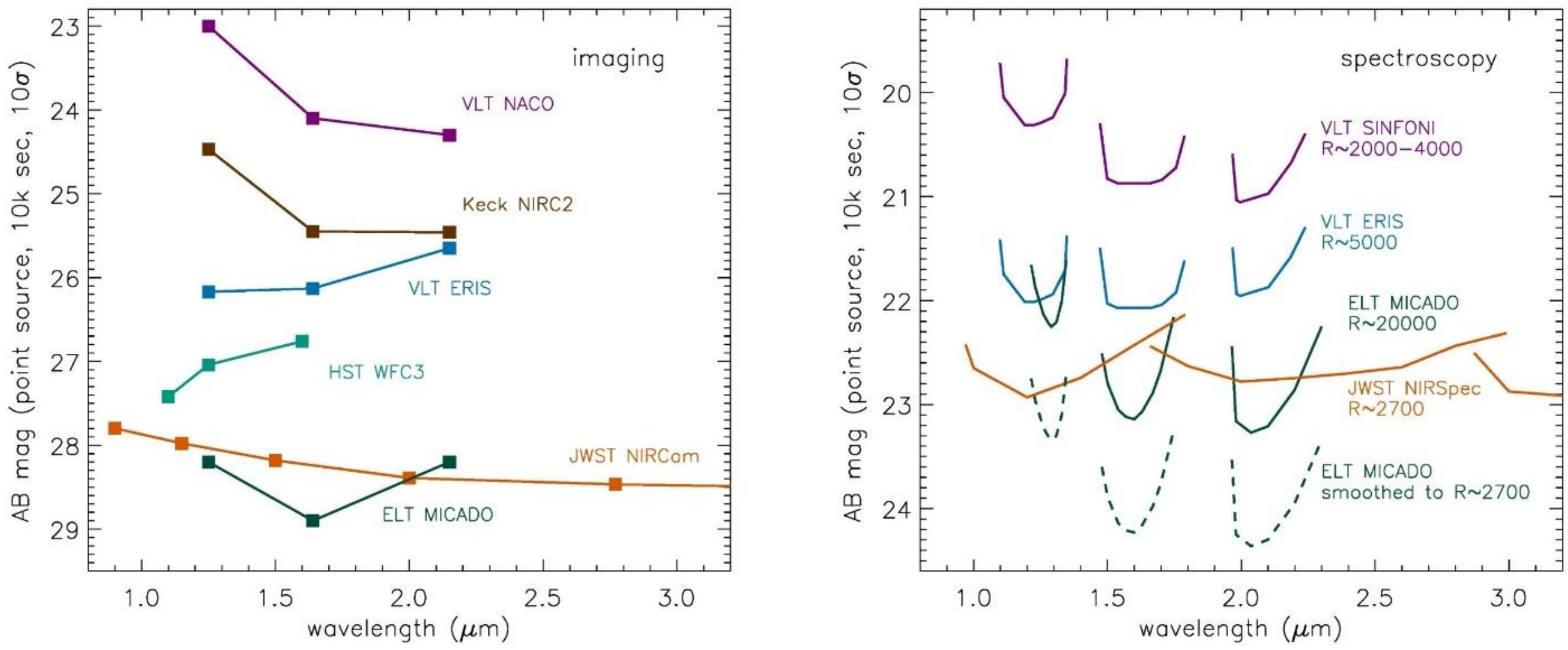

Figure 5: The sensitivity of MICADO in imaging (left) and spectroscopy (right) mode in the context of some of the main current telescopes and instruments.

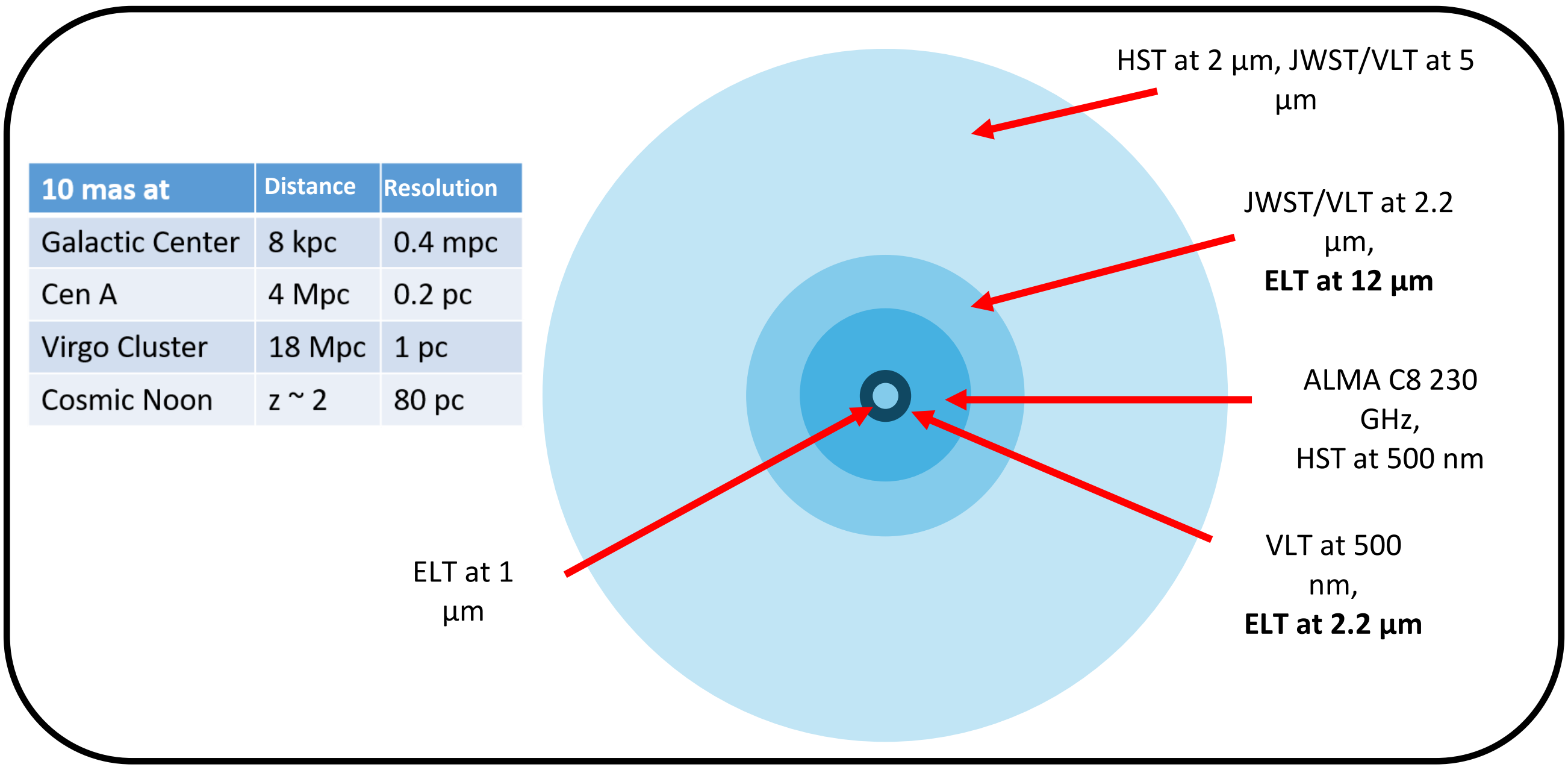

| 10 mas at | Distance | Resolution |
|---|---|---|
| Galactic Center | 8 kpc | 0.4 mpc |
| Cen A | 4 Mpc | 0.2 pc |
| Virgo Cluster | 18 Mpc | 1 pc |
| Cosmic Noon | z ~ 2 | 80 pc |

Figure 6: The spatial resolution of MICADO in context.

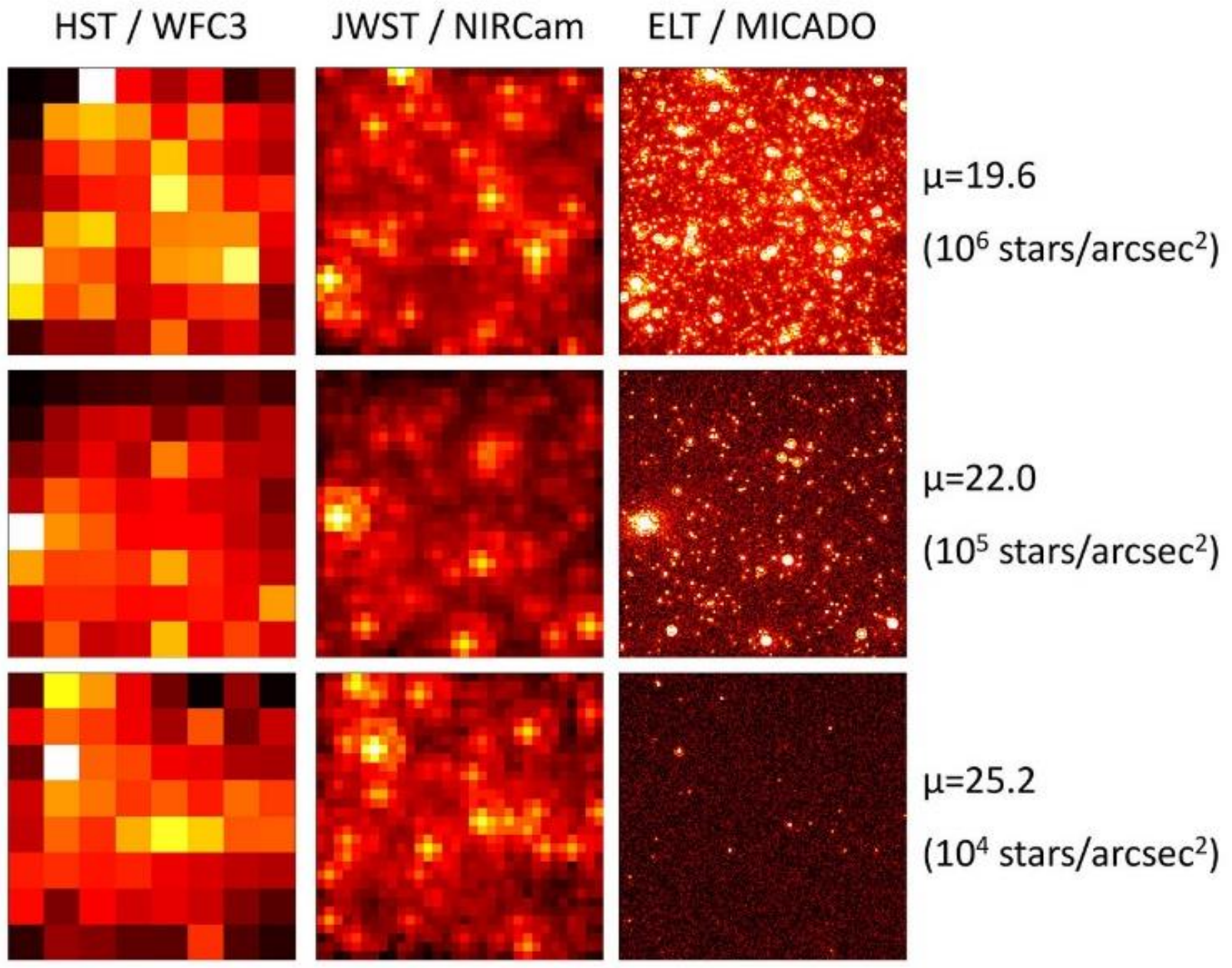

Figure 7: Simulations[1] of crowded stellar fields at various surface brightnesses, showing the impact of crowding on what can be measured by HST, JWST, and MICADO. The bottom row matches the stellar density at a radius of 4-5 $R_{eff}$ for NGC 4472 in the Virgo Cluster and represents the limit for JWST resolution. The top row corresponds to 2 $R_{eff}$ in the same galaxy and many individual stars can still be measured by MICADO. Each panel is 1” across.

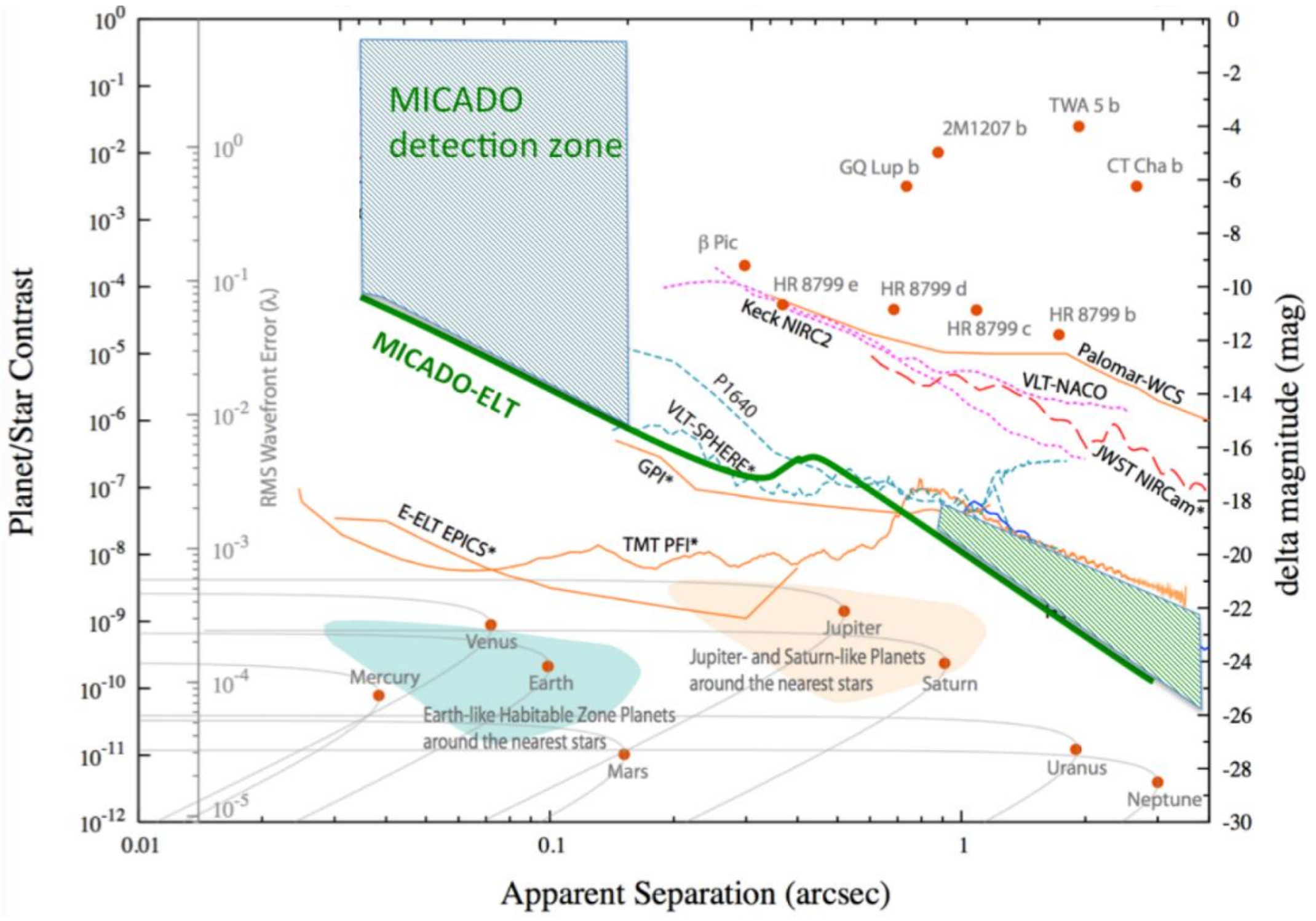

Figure 8: MICADO as an exoplanet camera[12].

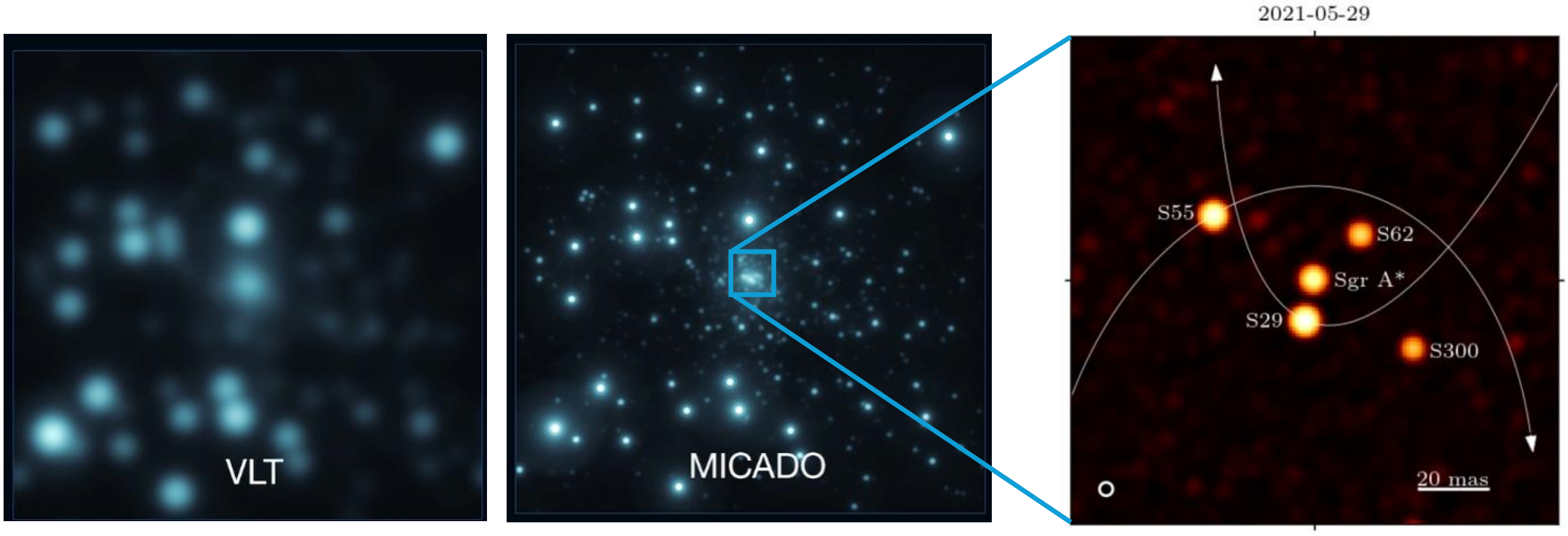

Figure 9: The Galactic Centre as seen by VLT/NACO (*left*)[13], VLTI/GRAVITY (*right*)[14] and a simulation for ELT/MICADO (*middle*).

With its classical coronagraph (occulting spot and Lyot stop, also available: vAPP and SAMs) and post processing (with ADI) MICADO can reach a contrast down to ~$10^{-6}$. In good seeing, an 800 K planet at 5 AU will be detectable in H & K-band in 30 mins. Thanks to the small inner working angle MICADO will characterise exoplanets in synergy with Gaia or RV surveys (Figure 8).

MICADO will be able to detect fainter main sequence stars with semi-major axes significantly smaller than the currently known S-stars, and determine their orbit (astrometry and Doppler spectroscopy, in combination with GRAVITY+). Such stars would have orbital time scales of a few years. In this case the Schwarzschild precession term and other GR terms will be detectable in a decade of observations (Figure 9).

## ACKNOWLEDGEMENTS

MICADO is being designed and built under the leadership of the Max Planck Institute for Extraterrestrial Physics (MPE) by a consortium of partners in Finland, Germany, France, the Netherlands, Austria and Italy, together with ESO. In addition to MPE, the MICADO consortium consists of the Max Planck Institute for Astronomy (MPIA, Germany), the University Observatory Munich (USM, Germany), the Institute for Astrophysics in Göttingen (IAG, Germany), NOVA (Netherlands Research School for Astronomy, represented by the University of Groningen, the University of Leiden, and the NOVA optical/infrared instrumentation group based at ASTRON), INAF (Italy), CNRS/INSU (the National institute For Earth Science and Astronomy of the French National Centre for Scientific Research, represented by the LESIA Space and Astrophysics Instrumentation Research Laboratory at Paris Observatory, the GEPI Galaxies, Stars, Physics and Instrumentation Laboratory at Paris Observatory, the UTINAM Universe, Time-Frequency, Interfaces, Nanostructures, Atmosphere-Environment and Molecules Institute, the LCF Charles Fabry Laboratory at Institut d'Optique Graduate School, and the IP2I Institute of Physics of 2 Infinites), A* (an Austrian partnership represented by the University of Vienna, the University of Innsbruck, the University of Linz, the Johann Radon Institute for Computational and Applied Mathematics at the University of Linz, and the Austrian Academy of Sciences), FINCA (the Finnish Centre for Astronomy with ESO), and the University of Turku (Finland). This work has benefited from the support of 1) the French Programme d'Investissement d'Avenir through the project F-CELT ANR-21-ESRE-0008, 2) the project WOLF ANR-18-CE31-0018, 3) the CNRS 80 PRIME program, 4) the CNRS INSU IR budget, 5) the Action Spécifique Haute Résolution Angulaire (ASHRA) of CNRS/INSU co-funded by CNES, 6), the Observatoire de Paris, 7) the Ile de France region (DIM ACAV/ACAV+ and ORIGINES), and 8) the INAF budget. USM is supported by the German Federal Ministry of Education and Research (BMBF) under the ErUM (Research of the Universe and Matter) grants 05A11WM1, 05A14WM1, 05A17WM1, 05A20WM1 and 05A23WM1.